\documentstyle[epsf,emulateapj]{article}
\newcommand{\ergps}{erg s$^{-1}$ }

\newcommand{\hmtwo}{h$_{\rm 50}^{-2}$ }
\newcommand{\hmthree}{h$_{\rm 50}^{-3}$ }
\begin{document}
\title{Discovery of a very X-ray luminous galaxy cluster
       at $z=0.89$ in the WARPS survey}
\author{H.\ Ebeling\altaffilmark{1,8}, L.R.\ Jones\altaffilmark{2,8},
        B.W.\ Fairley\altaffilmark{2},
        E.\ Perlman\altaffilmark{3,4}, C.\ Scharf\altaffilmark{5},
        D.\ Horner\altaffilmark{6,7}}
\altaffiltext{1}{Institute for Astronomy, 2680 Woodlawn Drive, Honolulu,
       Hawaii 96822, USA}
\altaffiltext{2}{School of Physics and Astronomy, University of
       Birmingham, Brimingham B15\,2TT, UK}
\altaffiltext{3}{Department of Physics and Astronomy, Johns Hopkins University,
	3400 North Charles Street, Baltimore, MD 21218, USA}
\altaffiltext{4}{Joint Center for Astrophysics, University of Maryland, Baltimore County, 1000 Hilltop Circle, Baltimore, MD 21250, USA}
\altaffiltext{5}{Space Telescope Science Institute, Baltimore, MD 21218, USA}
\altaffiltext{6}{Laboratory for High Energy Astrophysics, Code 660, NASA/GSFC, 
       Greenbelt, MD 20771, USA}
\altaffiltext{7}{University of Maryland, College Park, MD 20742-2421, USA}
\altaffiltext{8}{Visiting Astronomer at the W.M. Keck Observatory, jointly
                 operated by the California Institute of Technology and the
                 University of California.}

\slugcomment{accepted for publication in ApJL}

\begin{abstract}

We report the discovery of the galaxy cluster ClJ1226.9+3332 in the
Wide Angle ROSAT Pointed Survey (WARPS). At $z=0.888$ and $L_{\rm X} =
1.1\times 10^{45}$ \hmtwo \ergps ($0.5-2.0$ keV) ClJ1226.9+3332 is the
most distant X-ray luminous cluster currently known. The mere
existence of this system represents a huge problem for $\Omega_0=1$
world models.

At the modest (off-axis) resolution of the ROSAT PSPC observation in
which the system was detected, ClJ1226.9+3332 appears relaxed; an
off-axis HRI observation confirms this impression and rules out
significant contamination from point sources. However, in moderately
deep optical images (R and I band) the cluster exhibits signs of
substructure in its apparent galaxy distribution. A first crude
estimate of the velocity dispersion of the cluster galaxies based on
six redshifts yields a high value of 1650 km s$^{-1}$, indicative of a
very massive cluster and/or the presence of substructure along the
line of sight. While a more accurate assessment of the dynamical state
of this system requires much better data at both optical and X-ray
wavelengths, the high mass of the cluster has already been
unambiguously confirmed by a very strong detection of the
Sunyaev-Zel'dovich effect in its direction (Joy et al.\ 2001).

Using ClJ1226.9+3332 and ClJ0152.7--1357 ($z=0.835$), the second-most
distant X-ray luminous cluster currently known and also a WARPS
discovery, we obtain a first estimate of the cluster X-ray luminosity
function at $0.8<z<1.4$ and $L_{\rm X} > 5\times 10^{44}$ \hmtwo
\ergps ($0.5-2.0$ keV). Using the best currently available data, we
find the comoving space density of very distant, massive clusters to
be in excellent agreement with the value measured locally ($z<0.3$),
and conclude that negative evolution is not required at these
luminosities out to $z\sim 1$. Our findings are in conflict with
earlier claims of highly significant ($>3 \sigma$) negative evolution
already at $0.3<z<0.6$ based on the cluster samples of the EMSS and
the CfA 160 degree survey. Our results agree, however, with the lack
of significant evolution of very X-ray luminous clusters out to $z\sim
0.4$ reported by the MACS team. Our findings are also consistent with
the abundance of very X-ray luminous clusters at $z\sim 0.8$ inferred
from the EMSS cluster sample, provided ClJ0152.7--1357 (which was
missed by the EMSS) is added in.
\end{abstract}


\section{Introduction} 

Measurements of the abundance of clusters of galaxies as a function of
redshift allow a number of physical and cosmological parameters of
structure formation models to be constrained (e.g.\ Oukbir \&
Blanchard, 1997; Eke et al., 1998). The tightest constraints are
obtained from observations of the most massive and most distant
clusters which are extremely rare in all models of cluster formation.
For instance, the predicted space density of galaxy clusters with
intra-cluster gas temperatures of about $8\times 10^8$ K (k$T\sim 7$
keV) at $z\sim 1$ is two orders of magnitudes higher in an open
universe with $\Omega_0=0.3$ than in a flat universe with $\Omega_0=1$
(Viana \& Liddle, 1996).

Among the many efforts to compile statistically relevant samples of
these rare systems, X-ray flux limited surveys carry particular
appeal. Galaxy clusters are bright X-ray sources that can be detected
out to high redshift. Moreover, X-ray emission from clusters
originates from gas trapped and heated in deep gravitational potential
wells. X-ray surveys thus naturally select three-dimensionally bound
systems and are almost unaffected by projection effects. Finally,
X-ray flux limited surveys greatly facilitate the measurement of
comoving space densities since effective search volumes are easily
computable from the surveys' selection functions which are usually a
simple function of X-ray flux and, for distant clusters, almost
independent of X-ray source extent.

While one would ideally want to study clusters that are both distant
and massive, the rarity of these objects currently forces the observer
to give priority to one of the two qualifiers. Previous and ongoing
cluster surveys have thus adopted either of two fundamentally
different, but complementary, approaches.  For instance, very deep
X-ray and optical surveys covering only small areas of sky have become
increasingly successful at finding poor clusters at $z\ga 1$ (Rosati
et al., 1999; Lubin et al., 2000) while, at the other extreme,
relatively shallow X-ray surveys covering very large solid angles
(Ebeling, Edge \& Henry, 2001) are in the process of producing large
samples of very massive systems at lower redshift ($z\sim 0.5$).

\section{The Wide Angle ROSAT Pointed Survey}

The Wide Angle ROSAT Pointed Survey (WARPS) is one of a few surveys
straddling the dividing line between the two strategies outlined in
the previous section. Following the approach pioneered by the EINSTEIN
Extended Medium Sensitivity Survey (EMSS, Gioia et al., 1990; Stocke
et al., 1991), WARPS searches for distant clusters among a large
number of X-ray sources serendipitously detected in pointed ROSAT PSPC
observations. The WARPS strategy, as well as earlier results, have
been discussed in previous papers (Scharf et al., 1997; Jones et al.,
1998; Ebeling et al., 2000; Fairley et al., 2000).

\section{Discovery of ClJ1226.9+3332}

ClJ1226.9+3332 was detected in the WARPS survey as an extended X-ray
source 14.5 arcminutes off axis in the ROSAT PSPC observation of NGC
4395, a nearby, low-luminosity Sy1 galaxy (RP600277N00, on-axis
exposure time 9036s). The significance of detection exceeds 10
$\sigma$ in a 2 arcmin (radius) aperture. The source is also detected
at more than 5 $\sigma$ significance (within an aperture of 1' radius
and at an off-axis angle in 14.1') in a ROSAT HRI observation of the
same target (RH702725N00, on-axis exposure time 11,353s). 

Figure~1 shows an I band image of the source with adaptively smoothed
X-ray contours from the ROSAT PSPC and HRI observations overlaid.
The emission is well centered on an apparent overdensity of faint
galaxies.

\subsection{X-ray observations}
\label{xray}

The net PSPC count rate directly detected by the VTP algorithm
(Ebeling \& Wiedenmann, 1993; see also Scharf et al.\ 1997 for details
of how WARPS employs this detection algorithm) is $(2.29\pm
0.18)\times 10^{-2}$ ct s$^{-1}$ in PHA channels 50 to 200. The
exposure time at the location of the source ($\alpha=\rm 12^h\, 26^m\,
58.1^s, \delta= +33^{\circ}\, 32'\, 50.1''$, J2000) is 7.9 ks. Assuming a
beta model (Cavaliere \& Fusco-Femiano, 1976) with $\beta = 2/3$ and
core radius $70\,h_{\rm 50}^{-1}$ kpc, derived from the distribution
of the immediately detected photons (see Scharf et al.\ 1997 for
details), and taking into account the effects of the PSPC off-axis
point-spread function (PSF) we extrapolate to a total source count
rate of $(2.77\pm 0.21)\times 10^{-2}$ ct s$^{-1}$. Using the Galactic
value of $1.38\times 10^{20}$ cm$^{-2}$ for the equivalent column
density of neutral hydrogen in the direction of the source (Dickey \&
Lockman, 1990), a metallicity of 0.3, and a gas temperature of
$kT=11.7$ keV (see below) we convert the total PSPC count rate into a
total X-ray flux in the 0.5--2.0 keV band of $(3.4\pm 0.3)\times
10^{-13}$ \ergps cm$^{-2}$.

Following the same procedure for the HRI data, we find an exposure
time of 10.2 ks at the location of the source ($\alpha=\rm 12^h\,
26^m\, 58.0^s, \delta= +33^{\circ}\, 32'\, 54.1''$, J2000). This
source centroid agrees to within 4.2'' (less than the typical ROSAT
astrometry error) with the one determined from the PSPC data. The
extrapolated total cluster count rate is measured to be $(1.10\pm
0.18)\times 10^{-2}$ ct s$^{-1}$.  We use the same assumptions as
before to convert this count rate to a total X-ray flux of $(2.9\pm
0.2)\times 10^{-13}$ \ergps cm$^{-2}$ (0.5--2.0 keV), a value that is
in good agreement with the PSPC measurement. Although the HRI image
shows no obvious point sources within three arcminutes of the source
(see Fig.~1) we cannot strictly rule out the possibility that some or
all of the observed X-ray flux originates from a concentration of
individually faint point sources.

\subsection{Optical observations}

We obtained redshifts of 11 objects close to the apparent cluster core
shown in Fig.~1 using the LRIS spectrograph (Oke et al.~1995) on the
Keck-II 10m telescope.  Both apparent cluster members and possible
X-ray contaminants (blue galaxies and stars) were targeted. Combining
the results of a longslit observation carried out in January 1999
(300/5000 grism, dispersion 2.6\AA/pixel, slitwidth 1.5 arcsec,
wavelength range 5000--10000\AA) with those of multi-object
spectroscopy performed in January 2000 (600/7500 grism, dispersion
1.3\AA/pixel, slitwidth 1.5 arcsec, wavelength range typically
6200--8700\AA) we find no obvious contaminants (broad line emitters)
but six galaxies with accordant redshifts around a (heliocentric) mean
of $z=0.8877$. The six spectroscopically confirmed cluster members are
marked in Fig.~1; their positions and redshifts are listed in
Table~1. The position of the brightest cluster galaxy (labeled A in
Table~1) coincides within 2'' (6'') with the X-ray emission centroid
as determined from the PSPC (HRI) observations discussed in
Section~\ref{xray}, supporting our identification of this source as a
distant cluster of galaxies.

\section{Intrinsic properties of ClJ1226.9+3332}

From the six redshifts measured by us so far (see Table~1) we obtain a
very crude first estimate of the cluster galaxy velocity dispersion of
$\sigma = 1650^{+930}_{-340}$ km/s. If confirmed in future
observations, this very high value suggests a very massive cluster
and/or substructure along the line of sight.

Assuming a mean cluster redshift of $z=0.8877$ we derive a total X-ray
luminosity of $L_{\rm X} = (1.06\pm 0.08)\times 10^{45}$ \hmtwo \ergps
($0.5-2.0$ keV, $q_0=0.5$ is assumed throughout this Letter) from the
total cluster flux as determined from the PSPC observation,
corresponding to an estimated gas temperature of 11.7 keV based on the
$L_{\rm X}-kT$ relation of White, Jones \& Forman (1997). The inferred
bolometric luminosity of this sytem is $8.0^{+2.2}_{-1.9} \times
10^{45}$ \hmtwo \ergps (we assume an uncertainty of $\pm 1$ keV in the
estimated $kT$ value), making ClJ1226.9+3332 more X-ray luminous than
any other cluster currently known at $z>0.55$. The cluster's estimated
X-ray temperature is consistent with its high velocity dispersion
(Mushotzky \& Scharf, 1997).

While the X-ray and optical properties of ClJ1226.9+3332 described
above strongly suggest the presence of a highly massive cluster, the
limited depth of the existing X-ray and optical data does not allow us
to rule out contamination from a multitude of X-ray faint point
sources and/or from projection effects. Unambiguous evidence of the
high mass of this system was, however, obtained recently in the form
of the detection of a very strong Sunyaev-Zeldovich Effect (SZE)
decrement centered on the X-ray position reported here.  The SZE
measurements were obtained interferometrically at the
Berkeley-Illinois-Maryland Array (BIMA) at a frequency of 28.5 GHz,
and are reported by Joy et al.\ (2001).  The BIMA observations confirm
the presence of a deep gravitational potential well and lead to an
estimate for the total gravitational mass of ClJ1226.9+3332 that is
similar to, and possibly higher than that of the well studied, massive
cluster MS\,1054.4-0321 at $z=0.83$ (Hoekstra, Franx \& Kuijken 2000,
and references therein).

\section{The effective depth of serendipitous X-ray cluster surveys}

All serendipitous X-ray cluster surveys past and present, from the
EMSS to the ROSAT Deep Cluster Survey (RDCS, Rosati et al., 1998),
probe deep enough to detect very X-ray luminous clusters ($L_{\rm X}>8
\times 10^{44}$ \hmtwo \ergps, 0.5--2.0 keV) out to redshifts $z\ga
1.5$. WARPS, for instance, would have detected ClJ1226.9+3332 over the
full geometric solid angle covered (72.0 deg$^2$) out to a maximal
redshift of $z=1.87$. 

Although formally correct, such maximal detection redshifts are
misleading. WARPS, like all other serendipitous cluster surveys
conducted to date (with the notable exception of the RDCS), performed
its imaging follow-up observations exclusively in the optical R and I
bands. For clusters at $z>1$, R and I correspond to B or even U in the
cluster rest frame and are thus very inefficient bands for the
detection of distant cluster ellipticals. Indeed, and not
surprisingly, WARPS\footnote{as well as all other cluster surveys
relying exclusively on imaging observations in the optical passband
for cluster identifications} failed to detect any cluster at $z>1$. We
thus argue that the limitations of the optical imaging observations
impose a more stringent constraint of $z_{\rm max}\sim 1$.

Imaging of X-ray selected cluster candidates at near-infrared
wavelengths (as conducted by the RDCS team) allows the identification
of clusters beyond $z\sim 1$ (Rosati et al., 1999).  However,
instrumental limitations still impose a redshift limit of $z\sim 1.4$
as, at $z>1.4$, the most important absorption features (the Ca H and K
doublet) are redshifted to observed wavelengths redward of 9,500 \AA\
where the efficiency of even the most powerful present-day
spectrograph (LRIS on the Keck 10m telescope) is less than 5 per
cent. Consequently even the currently deepest X-ray cluster survey,
the RDCS, has yet to spectroscopically confirm a cluster at $z>1.3$.

\section{The evolution of the most massive clusters of galaxies out to
         $z\sim 1$}

We attempt to constrain the evolution of the X-ray cluster luminosity
function (XLF) at the highest redshifts and highest luminosities using
the small, but statistically complete WARPS sample defined by the
selection criteria $z>0.8$ and $L_{\rm X} > 5\times 10^{44}$ \hmtwo
\ergps ($0.5-2.0$ keV). The only two WARPS clusters meeting these
criteria, ClJ0152.7--1357 (Della Ceca et al., 2000; Ebeling et al.,
2000) and ClJ1226.9+3332, are the most X-ray luminous distant clusters
currently known\footnote{A discussion of the WARPS XLF of less
luminous systems at $z>0.8$ will be presented elsewhere.}.  Both
ClJ0152.7--1357 (which was missed by the EMSS, Ebeling et al., 2000)
and ClJ1226.9+3332 feature X-ray fluxes well above the WARPS flux
limit and would have been detected out to maximal redshifts of
$z=1.57$ and $z=1.87$, respectively. However, for the reasons
described in the previous section, we use a lower maximal redshift of
$z=1$ when computing the WARPS search volumes.

The resulting cumulative WARPS X-ray luminosity function of very X-ray
luminous ($L_{\rm X} > 5\times 10^{44}$ \hmtwo \ergps, $0.5-2.0$ keV)
and very distant ($z>0.8$) clusters is shown in Figure~2. Standard
$1\sigma$ errors (in the Poisson limit, Gehrels 1986) are indicated by
the dark, shaded region. We also compute an alternative version of the XLF,
using the very generous assumption of $z_{\rm max} = 1.4$ (shown by
the hatched region in Fig.~2). The currently best determination of the
local ($z<0.3$) cluster XLF from the ROSAT Brightest Cluster Sample
(Ebeling et al., 1997) is shown by the thick, solid line. We find both
WARPS XLF estimates to be in very good agreement with the local
measurement. Since the search volumes of WARPS and EMSS for clusters
with $L_{\rm X}\sim 9\times 10^{44}$ \hmtwo \ergps ($0.5-2.0$ keV) at
$0.8<z<1$ are very similar (about $1\times 10^{8}$ \hmthree Mpc$^{3}$)
our measurement is also in good agreement with the XLF inferred from
the two X-ray luminous EMSS clusters in this redshift interval,
MS1054--0321 ($z=0.829$) and ClJ0152.7--1357 (originally missed by the
EMSS).

Also shown in Fig.~2 is the prediction of the XLF evolution model
derived by Rosati et al.\ (2000) from a fit to the RDCS data, together
with its $2\sigma$ errror range. At $z\sim 0.9$ this model lies
between a factor of 100 and a factor of 1000 below the WARPS XLF
measurement. While our data thus do not support the RDCS model, they
cannot strictly rule it out either because of the large uncertainties
in the model parameters. We note though that at $L_{\rm X}=5-11\times
10^{44}$ \hmtwo \ergps ($0.5-2.0$ keV) the observed WARPS XLF for very
X-ray luminous clusters at $0.8\le z\le 1$ is inconsistent with the
RDCS prediction at greater than $2\sigma$ significance.

\section{Conclusions}

The discovery of ClJ1226.9+3332, the most X-ray luminous distant
cluster currently known, adds to the growing evidence in favour of an
early period of cluster formation at redshifts $z\ga 1$ with little
evolution in the cluster abundance ever since.

While the small size of the WARPS subsample discussed here does not
allow us to firmly rule out negative or positive evolution at $z\sim
1$, we stress that our data certainly do not {\em require}\/ a change
in the abundance of very X-ray luminous clusters out to the highest
redshifts probed by current surveys.  Specifically, we do not find a
decrease in the comoving cluster abundance by a factor of more than 100
(at $z\sim 0.9$ compared to the local value) as predicted by the
best-fitting XLF model derived from RDCS data (Rosati et al., 2000).

The lack of significant evolution observed by us at the highest
redshifts and highest X-ray luminosities is consistent with the EMSS
results at $z>0.8$ if the known incompleteness of the EMSS at these
redshifts is corrected for.  Our findings also agree with preliminary
results from the MAssive Cluster Survey (MACS, Ebeling, Edge \& Henry,
2000) which, boasting greatly improved statistics, finds only very
mild negative evolution in a measurement of the XLF of similarly X-ray
luminous clusters at $z\sim 0.4$.

Unless cluster evolution is a non-monotonic function of redshift
and/or X-ray luminosity, our results are in conflict with earlier
claims of highly significant ($>3\sigma$) negative evolution at lower
redshifts ($0.3\le z\le 0.6$) and lower luminosities ($L_{\rm X} \ga
3\times 10^{44}$ \hmtwo \ergps, $0.5-2.0$ keV) based on the cluster
samples compiled in the EMSS (Gioia et al., 1990) and the CfA 160 deg
survey (Vikhlinin et al., 1998a).

Larger, well defined samples of massive clusters at intermediate and
high redshift as well as spatially resolved cluster temperatures are
needed to actually measure the rate of evolution and to place
meaningful constraints on the cosmological parameters governing
structure formation.

\acknowledgements

We thank the telescope time allocation committee of the University of
Hawai`i for their generous support of the WARPS optical follow-up
programme. HE gratefully acknowledges financial support from NASA LTSA
grant NAG 5-8253. LRJ thanks the UK PPARC for financial support.

\begin{deluxetable}{llllll} 
\tablecolumns{4}
\tablewidth{0pc} 
\tablehead{
\colhead{galaxy} &
\colhead{R.A. (J2000)} &
\colhead{Dec (J2000)} &
\colhead{redshift $z$}} 
\tablecaption{Positions and heliocentric redshifts of confirmed cluster members}
\tablecomments{The quoted positions are accurate to better than 1 arcsec.
              All redshifts were derived from spectra taken with LRIS on
              Keck-II. The quoted redshift errors are $1\sigma$ standard
              deviations from the quoted mean and are based on the observed wavelengths of
              individual spectral features.}
\startdata 
A   & $12^h\, 26^m\, 58.2^s$ & $+33^{\circ}\, 32'\, 49''$ & $0.8911 \pm 0.0001$  \\
B   & $12^h\, 26^m\, 57.5^s$ & $+33^{\circ}\, 33'\, 14''$ & $0.8794 \pm 0.0004$  \\
C   & $12^h\, 26^m\, 56.1^s$ & $+33^{\circ}\, 32'\, 23''$ & $0.8985 \pm 0.0005$  \\
D   & $12^h\, 26^m\, 55.6^s$ & $+33^{\circ}\, 32'\, 13''$ & $0.9006 \pm 0.0005$  \\
E   & $12^h\, 26^m\, 53.0^s$ & $+33^{\circ}\, 32'\, 50''$ & $0.8798 \pm 0.0003$  \\
F   & $12^h\, 26^m\, 50.6^s$ & $+33^{\circ}\, 32'\, 46''$ & $0.8769 \pm 0.0004$  \\
\enddata
\end{deluxetable}

\begin{figure}
\epsfxsize=\textwidth
\epsffile{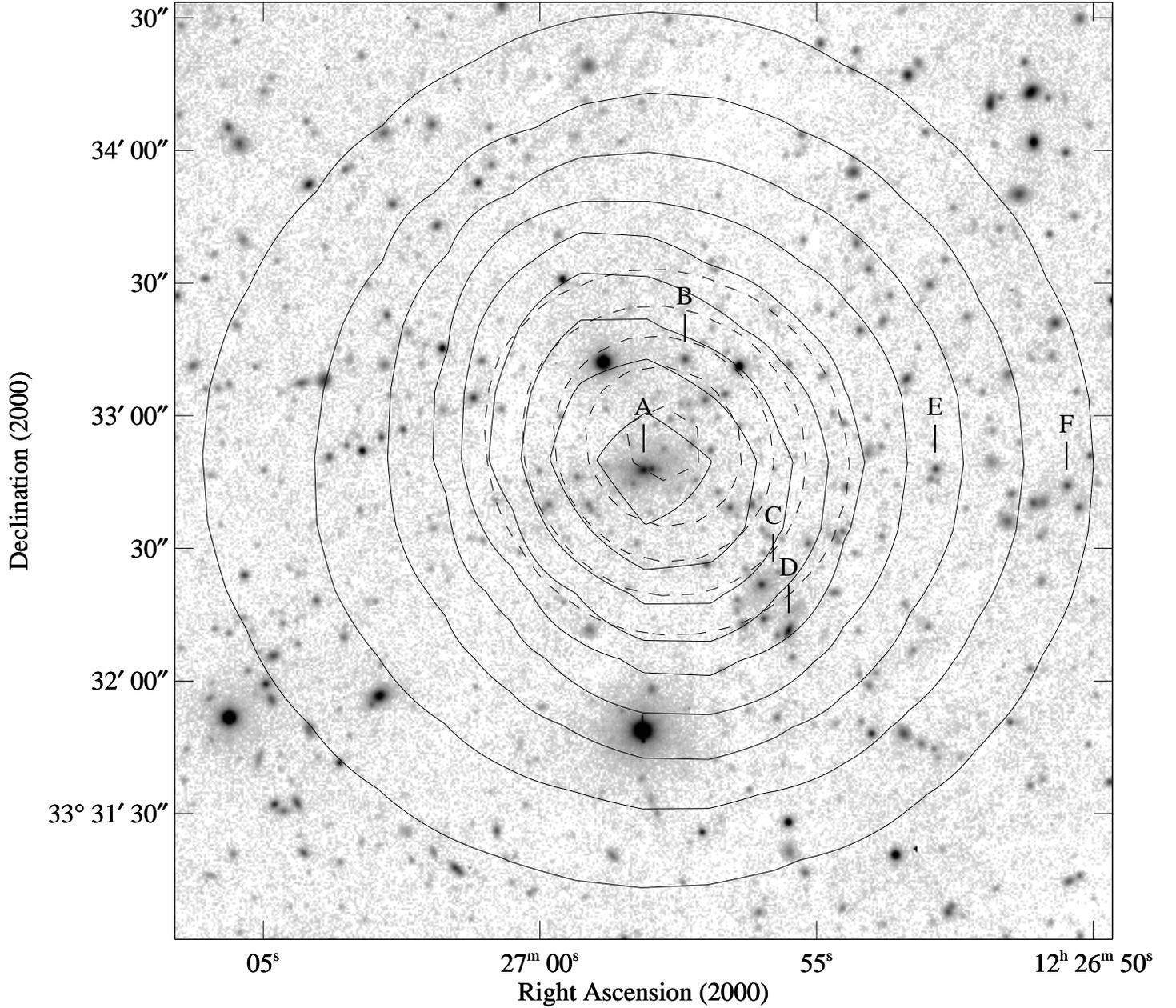}
\caption{I band image of ClJ1226.9+3332 obtained in a four minute
exposure with LRIS on Keck-II in February 1999. Overlaid are
logarithmically spaced contours of the adaptively smoothed X-ray
emission as seen with the ROSAT PSPC in the 0.5--2.0 keV band (solid
lines), and as seen with the ROSAT HRI (dashed lines). The lowest PSPC
(HRI) contour lies a factor of 2.5 (1.2) above the background value of
$2.5\times 10^{-4}$ ($5.4\times 10^{-3}$) ct s$^{-1}$ arcmin$^{-2}$;
the levels of adjacent contours differ by a factor of 1.5 (1.2).  The
FWHM of the PSPC (HRI) point-spread function at this off-axis angle
(14 arcmin) is 37 (6) arcsec.}
\end{figure}

\begin{figure}
\epsfxsize=0.9\textwidth
\epsffile{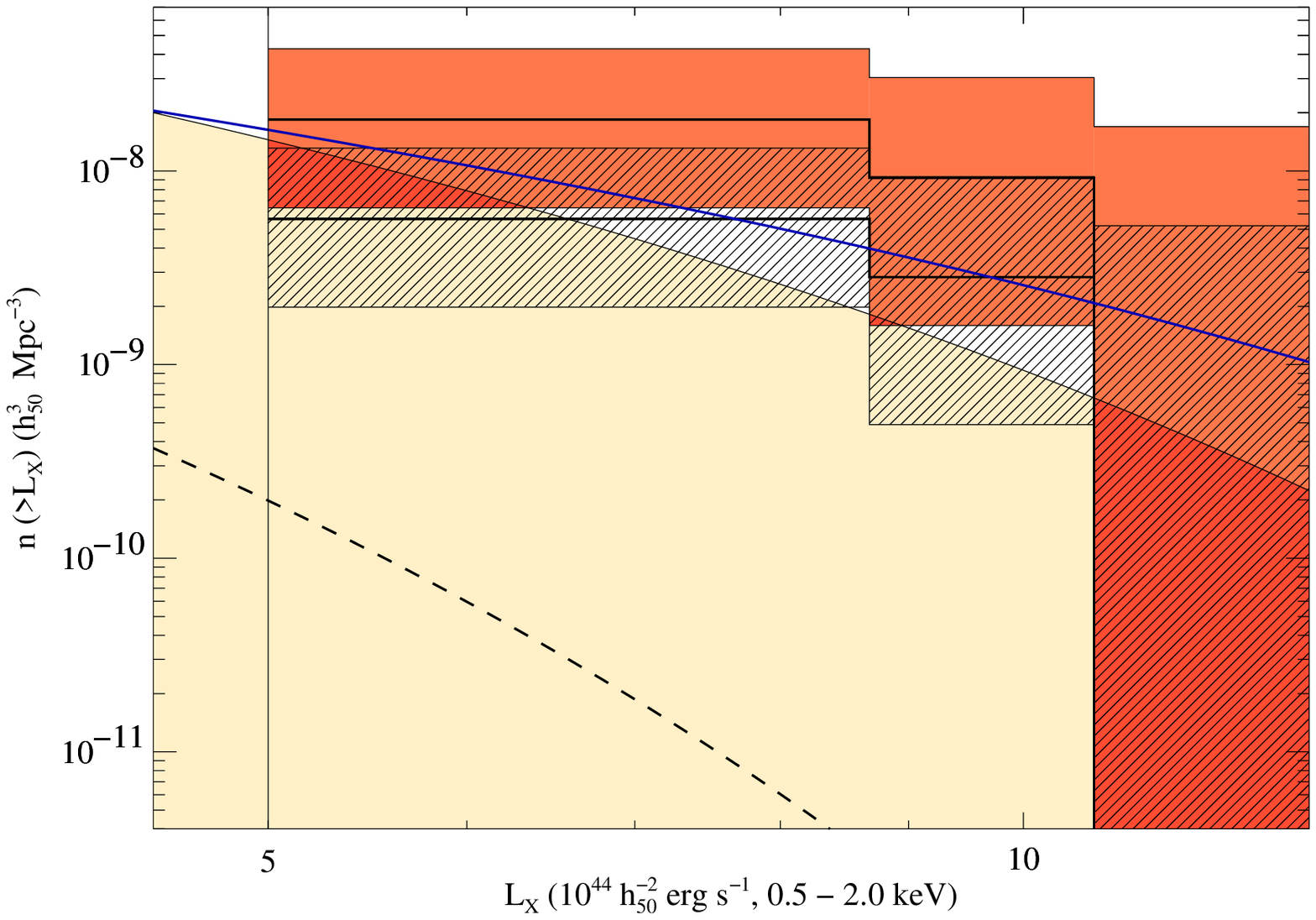}
\caption{The cumulative X-ray luminosity function (XLF) of very X-ray
luminous, distant ($z>0.8$) clusters based on ClJ0152.7--1357 and
ClJ1226.9+3332 as detected in the WARPS survey (solid lines). The
strongly shaded region shows the $1\sigma$ Poisson error range around
the XLF assuming a maximal detection redshift of $z=1$; the hatched
region just below delineates the $1\sigma$ Poisson error range for the
WARPS XLF assuming a maximal detection redshift of $z=1.4$. The local XLF
based on the ROSAT Brightest Cluster Sample (Ebeling et al., 1997) is
shown by the bold, solid line, whereas the prediction of the RDCS XLF
evolution model (Rosati et al.\ 2000) with its $2\sigma$ error range
is shown by the dashed line and the lightly shaded region.}
\end{figure}


\begin{references}
\reference{beta} Cavaliere, A.\ \& Fusco-Femiano, R. 1976, A\&A, 49, 137
\reference{rdcs} Della Ceca R., Scaramella R., Gioia I.M., Rosati P., Fiore F.,
	Squires G. 2000, A\&A, 353, 498
\reference{nh} Dickey, J.M.\ \& Lockman, F.J. 1990, Ann.\ Rev.\ Astron.\
               Astroph., 28, 215
\reference{bcs} Ebeling H., Edge A.C., Fabian A.C., Allen S.W., 
	Crawford C.S., B\"ohringer H. 1997, ApJ, 479, L101
\reference{macs-1} Ebeling H., Edge A.C.\ \& Henry J.P. 2000, in {\em
 Large-Scale Structure in the X-ray Universe}, eds Plionis \& Kolokotronis,
        Atlantisciences, p39
\reference{macs-2} Ebeling H., Edge A.C.\ \& Henry J.P. 2001, ApJ, submitted
\reference{warps-3} Ebeling H.\ et al. 2000, ApJ, 534, 133
\reference{bruce} Fairley B.W., Jones L.R., Scharf C., Ebeling H., Perlman E.,
     	Horner D., Wegner G., Malkan M. 2000, MNRAS, 318, 333
\reference{gehrels} Gehrels N. 1986, \apj, 303, 336
\reference{gioia} Gioia I.M., Maccacaro T., Schild R.E., Wolter A.,
	Stocke J.T., Morris S.L., Henry J.P. 1990, ApJS, 72, 567
\reference{emss} Gioia I.M., Henry J.P., Maccacaro T., Morris S.L., 
	Stocke J.T., Wolter A. 1990, ApJ, 356, L35
\reference{hoekstra} Hoekstra, H., Franx, M., Kuijken, K. 2000, ApJ, 532, 88
\reference{jones} Jones L.R., Scharf C., Ebeling H., Perlman E., Wegner G., 
        Malkan M., Horner D. 1998, \apj, 495, 100
\reference{joy} Joy M.\ et al. 2001, ApJL, submitted, astro-ph/0012052
\reference{lori} Lubin L.M. Brunner R., Metzger M.R., Postman M., Oke J.B. 
 	2000, ApJ, 531, L5
\reference{kts} Mushotzky R.F.\ \& Scharf C.A. 1997, ApJ, 482, L13
\reference{ob2} Oukbir J.\ \& Blanchard A. 1997, A\&A, 317, 10
\reference{piero2} Rosati P., Della Ceca R., Norman C., Giacconi R. 1998, 
        \apj, 492, L21 
\reference{piero3} Rosati P., Stanford S.A., Eisenhardt P.R., Elston R., 
	Spinrad H., Stern D., Dey A. 1999, AJ, 118, 76
\reference{piero4} Rosati P., Borgani S., Della Ceca R., Stanford
	S.A., Eisenhardt P.R., Lidman C. 2000, in {\em Large-Scale
	Structure in the X-ray Universe}, eds Plionis \& Kolokotronis,
        Atlantisciences, p13
\reference{caleb} Scharf C.A., Jones L.R., Ebeling H., Perlman E., 
        Malkan M., Wegner G. 1997, \apj, 477, 79, Paper I
\reference{john} Stocke J.T., Morris S.L., Gioia I.M., Maccacaro T.,
	Schild R., Wolter A. 1991, ApJS, 76, 813
\reference{viana} Viana P.T.P.\ \& Liddle A.R. 1996, MNRAS, 281, 323
\reference{alexey1} Vikhlinin A., McNamara B.R., Forman W., Jones C., 
     	Quintana H., Hornstrup A., 1998a, ApJ, 498, L21
\reference{alexey2} Vikhlinin A., McNamara B.R., Forman W., Jones C., 
     	Quintana H., Hornstrup A., 1998b, ApJ, 502, 558
\reference{dave} White D.A., Jones C.\ \& Forman W. 1997, MNRAS, 292, 419
\end{references}
\end{document}